\documentclass{article}
\usepackage{amsfonts}
\usepackage{mathtools}
\usepackage{spconf,amsmath,graphicx,epsfig}
\usepackage{subcaption}
\usepackage{mwe}
\usepackage{xcolor}
\usepackage{hyperref}
\usepackage{algorithm}
\usepackage{algorithmic}
\usepackage{eqnarray}
\usepackage{adjustbox}
\usepackage{booktabs}
\usepackage{setspace}
\usepackage{verbatim}

\title{Neural Model Reprogramming with Similarity Based Mapping for Low-Resource Spoken Command Recognition}
\name{%
\begin{tabular}{@{}c@{}}
Hao Yen$^{1,2}$ \qquad
Pin-Jui Ku$^{1,2}$ \qquad
Chao-Han Huck Yang$^{1}$\qquad Hu Hu$^{1}$ \\ Sabato Marco Siniscalchi$^{1,3}$\qquad Pin-Yu Chen$^{4}$\qquad Yu Tsao$^{2}$ 
\end{tabular}}
\address{$^1$Georgia Institute of Technology, USA\\$^2$Research Center for Information Technology Innovation, Academia Sinica, Taiwan \\ $^3$Kore University of Enna, Italy and $^4$IBM Research, USA}

\begin{document}
\ninept
\maketitle

%

\begin{abstract}
In this study, we propose a novel adversarial reprogramming (AR) approach for low-resource spoken command recognition (SCR) and build an AR-SCR system. The AR procedure aims at repurposing a pretrained SCR model (from the source domain) to modify the acoustic signals (from the target domain). To solve the label mismatches between source and target domains and further improve the stability of AR, we propose a novel similarity-based label mapping technique to align classes. In addition, the transfer learning (TL) technique is combined with the original AR process to improve the model adaptation capability. We evaluate the proposed AR-SCR system on three low-resource SCR datasets, including Arabic, Lithuanian, and dysarthric Mandarin speech. Experimental results show that with a pretrained AM trained on a large-scale English dataset, the proposed AR-SCR system outperforms the current \textbf{state-of-the-art results} on Lithuanian and Arabic speech commands datasets, with only a limited amount of training data.

\end{abstract}

\section{Introduction}

The aim of spoken command recognition (SCR) is to identify a target command out of a set of predefined candidates, based on an input utterance~\cite {Zeppenfeld1992, Rohlicek1989}. Owing to its wide applicability to various domains, such as smart home devices~\cite{Bahpai2019} or crime defections~\cite{Kavya2014}, SCR has long been an important research topic in the speech processing field~\cite{Dinushika2019, Seo2021}. Due to recent advances in deep learning (DL) algorithms, the performance of SCR systems has been significantly enhanced~\cite{Andrade2018, majumdar2020matchboxnet, vygon2021learning}. However, a common requirement to build a high-performance DL-based SCR system is to prepare a large amount of labeled training data (speech utterances and corresponding transcriptions of the commands). Such a requirement is not always realizable in real-world scenarios. In fact, it is generally favorable to build an SCR system with only a limited amount of training data. Such a scenario is often referred to as a low-resource training scenario~\cite{Besacier2014}. 

Numerous algorithms have been developed to train DL-based systems under low-resource scenarios. A well-known category of approaches is transfer learning (TL)~\cite{Pan2010}, which aims to use a small amount of training data to fine-tune a pretrained model, where the pretrained model is generally trained on a large-scale dataset. Prior arts have demonstrated the effectiveness of TL in speech processing tasks. For example, in~\cite{Tetariy2015}, an English acoustic model (AM) pretrained on a large-scale training set is fine-tuned with limited training data to obtain a Spanish AM. Meanwhile in~\cite{menon2019feature}, a multilingual bottleneck feature extractor is pretrained on a large-scale training set and fine-tuned to form a keyword recognizer on a low-resourced Luganda corpus. The study of~\cite{lin2020training} collected 200 million pieces of 2-second data from Youtube to pretrain a speech embedding model to extract useful features for a downstream keyword spotting task. Although these TL approaches show promising results, the fine-tuning process (which is often done in the online mode) requires large training resources and thus is only feasible for applications where sufficient computation resources are available.  

Another category of approaches is to adopt a pretrained model to extract representative features to facilitate efficient and effective training for the SCR systems. The pretrained model is generally trained on a large-scale dataset with either a supervised or self-supervised training manner. In~\cite{mcmahan2018listening} and~\cite{myer2018efficient}, the SCR systems were established by adopting representative features extracted from a pretrained sound event detector and phone classifier, respectively. Both were trained on large-scale labeled data. Meanwhile, several methods adopt self-supervised models, such as PASE+~\cite{pascual2019learning, ravanelli2020multi}
, wav2vec~\cite{Schneider2019wav2vecUP, baevski2019vq}, and wav2vec 2.0~\cite{baevski2020wav2vec}, as feature extractors to build the SCR systems~\cite{mittal2021representation, Kolesau2020, Seo2021}. A notable drawback of this category of approaches is that an additional large-scale DL-based model (accordingly with increased hardware) is required.

\begin{figure}[t]
    \centering
    \includegraphics[width=\linewidth]{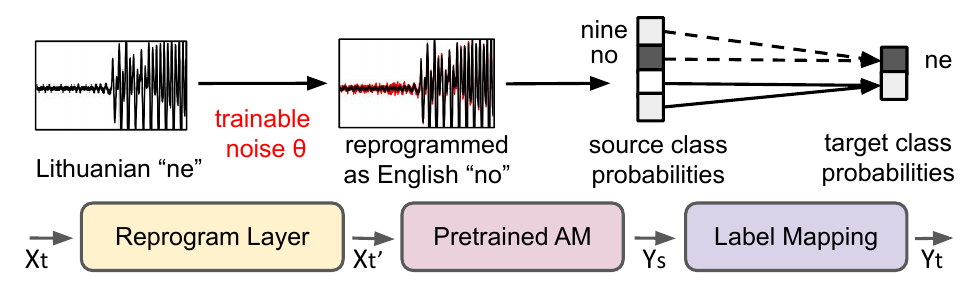}
    \caption{Illustration of the proposed AR-SCR system. 
    The acoustic signals of a Lithuanian command (``ne'') is reprogrammed to English commands (``nine'' and ``no'') and mapped to its final prediction with a pretrained English acoustic model. }
    \label{fig:illustration}
\vspace{-0.4cm}
\end{figure}

Adversarial reprogramming (AR), as an alternative model adaptation technique, has been confirmed to provide satisfactory results in numerous machine learning tasks ~\cite{Elsayed2018}. In~\cite{pmlr-v139-yang21j}, AR adopts a trainable layer to generate additive noise (e.g., as additional information) on input acoustic signals (source-domain data) to guide an AM to recognize electrocardiography (ECG) signals (target-domain data). Along with the success of~\cite{pmlr-v139-yang21j}, our study investigates whether AR can be applied to domain adaptation and accordingly building an SCR system in a low-resource training scenario. Different from ~\cite{Neto1995, Yao2012}, which perform acoustic model adaptation by passing input through a neural network layer for transformation, AR aims to add a trainable noise directly to the input sequence without modifying the model. Fig.~\ref{fig:illustration} shows the design concept of the proposed AR-SCR system, which consists of a reprogram layer and a pretrained AM. The reprogram layer first generates trainable noises, $\mathbf{\theta}$, to modify the original signals before passing them into the pretrained AM. The AM will then output class probabilities corresponding to the source classes. A label mapping technique is adopted to map the probabilities of source classes to the target class by aggregating probabilities over the assigned source labels. Based on the aggregated probabilities, the reprogram layer is further trained to generate noises and thus modify the input signals, so that the pretrained AM can be repurposed to perform recognition in the target task. 

The proposed AR-SCR system adopts two additional techniques to further improve the model adaptation capability: (1) a novel similarity-based label mapping strategy that aims to align the target and source classes and (2) a fine-tuning process that adjusts the AM with the AR-generated signals. Experiment results on three low-resource SCR datasets, including Arabic, Lithuanian, and dysarthric Mandarin speech command datasets, demonstrate that the proposed AR-SCR system can yield better performance than other state-of-the-art methods. In summary, the major contribution of the present work is twofold: 
1) This is the first study that investigates the applicability of AR to low-resource SCR tasks with promising results.
2) We verify that AR has the flexibility to combine with the TL technique to achieve a better model adaptation performance.  

\begin{figure}[t]
    \centering
    \includegraphics[width=\linewidth]{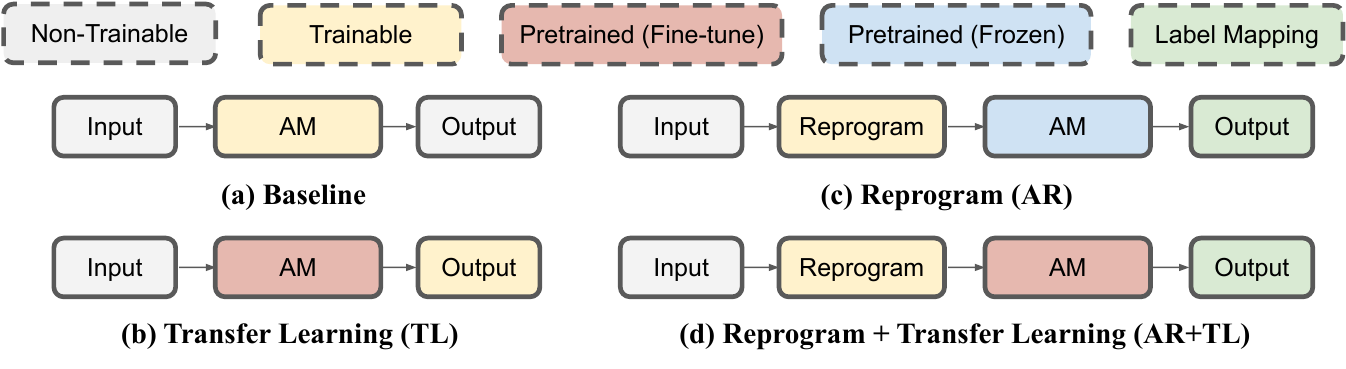}
    \caption{Frameworks studied in this work. The "AM" block refers to the acoustic model. In (a), the baseline system is trained from scratch on the target domain data. In (b), AM is pretrained and then fine-tuned on the target domain data. In (c), AM is pretrained on the source domain and then fixed; an adversarial reprogram (AR) layer is then placed before the pretrained AM model to modify the input signals. In (d), we combine AR and TL to train the adversarial reprogram layer and fine-tune AM simultaneously.}
    \label{fig:framework}
\end{figure}

\section{The Proposed AR-SCR System}
\label{sec:system}
\subsection{AR-SCR System}
Fig.~\ref{fig:framework} illustrates the overflow of an SCR system with AR and TL model adaptation techniques. In Fig.~\ref{fig:framework} (a), the AM is directly trained by data from the target-domain task. When the training data from the target domain is limited, the model cannot be trained well and thus may result in unsatisfactory recognition performance. In Fig.~\ref{fig:framework} (b), the TL technique is applied to the pretrained AM to establish a new SCR system that matches the target domain. In Fig.~\ref{fig:framework} (c), the pretrained AM is fixed, and an adversarial reprogram layer is trained to transform the input signals to reduce the distance between the target and source distributions. In Fig.~\ref{fig:framework} (d), the adversarial reprogram layer is treated as a front-end processor, and the TL technique is applied to further fine-tune AM with the reprogrammed signals. We expect the combination of AR (as a front-end processing) and TL (as a back-end processing) can reach better model adaptation capability due to their complementary abilities.

\subsection{Acoustic Signal Reprogramming}
The concept of AR was first introduced in~\cite{Elsayed2018}, and its aim was to determine a trainable input transformation function $\mathcal{H}$ to repurpose a pretrained model from the source domain to a target task. The authors in~\cite{Elsayed2018} showed that by the AR process, a pretrained ImageNet model trained on the image classification task can solve a square-counting task with high accuracy. 
A later study~\cite{Tsai2020} demonstrated that a reliable classification system can be established using AR and a black-box pretrained model with scarce data and limited resources. Meanwhile, the Voice2Series method~\cite{pmlr-v139-yang21j} is proposed to transfer time series data (e.g., ECG or Earthquake) $x_{t}$, as the target domain, $\mathcal{X}_{\mathcal{T}} \subseteq \mathbb{R}^{d_{T}}$ from the source domain $\mathcal{X}_{\mathcal{S}} \subseteq \mathbb{R}^{d_{s}}$, where $d_{\mathcal{T}}<d_{\mathcal{S}}$. For these AR approaches, a reprogrammed sample $x_{t}^\prime$ can be formulated as:
\begin{equation}
x_{t}^{\prime}=\mathcal{H}\left(x_{t} ; \theta\right):=\operatorname{Pad}\left(x_{t}\right)+\underbrace{M \odot \theta}_{\delta},
\label{eqn:orig_repr}
\end{equation}
where $\operatorname{Pad}\left(x_{t}\right)$ generates a zero-padded time series of dimension $d_{\mathcal{S}}$. The binary mask $M \in\{0,1\}^{d_{\mathcal{S}}}$ indicates the indexes that are not occupied and reprogrammable. $\theta \in \mathbb{R}^{d_{\mathcal{S}}}$ is a set of trainable parameters for aligning source and target domain data distributions. The term $\delta$ denotes the trainable additive input transformation for reprogramming. In the original AR method, the target sequence must be shorter than the source counterpart. To overcome this limitation, we design to add trainable noises to the whole sequence, and thus Eq.~\eqref{eqn:orig_repr} becomes
\begin{equation}
x_{t}^{\prime}=\mathcal{H}\left(x_{t} ; \theta\right):=x_{t}+\theta.
\label{eqn:repr}
\end{equation}

In this work, we focus on applying AR as a model adaptation technique to effectively fine-tune a pretrained model. 

\subsection{Pretrained Acoustic Model}
In the Voice2Series study~\cite{pmlr-v139-yang21j}, the authors have compared several well-known AMs as pretrained models and provided the first theoretical justifications by optimal transport for reprogramming general time-series signals to acoustic signals. Based on the provided justifications, in this study, we established AM with two layers of fully convolutional neural networks, followed by two bidirectional recurrent neural networks, which are then combined with an attention layer. To train the model, we use the Google Speech Commands dataset~\cite{Warden2018}, which is a large-scale collection of spoken command words, containing 105,829 utterances of 35 words from 2,618 speakers; all the utterances are recorded in a 16 kHz sampling-rate format. The pretrained AM has 0.2M parameters and yields a 96.90\% recognition accuracy rate on the testing set of Google Speech Commands.

\subsection{Similarity Label Mapping}
\begin{figure}[t]
    \centering
    \includegraphics[width=\linewidth]{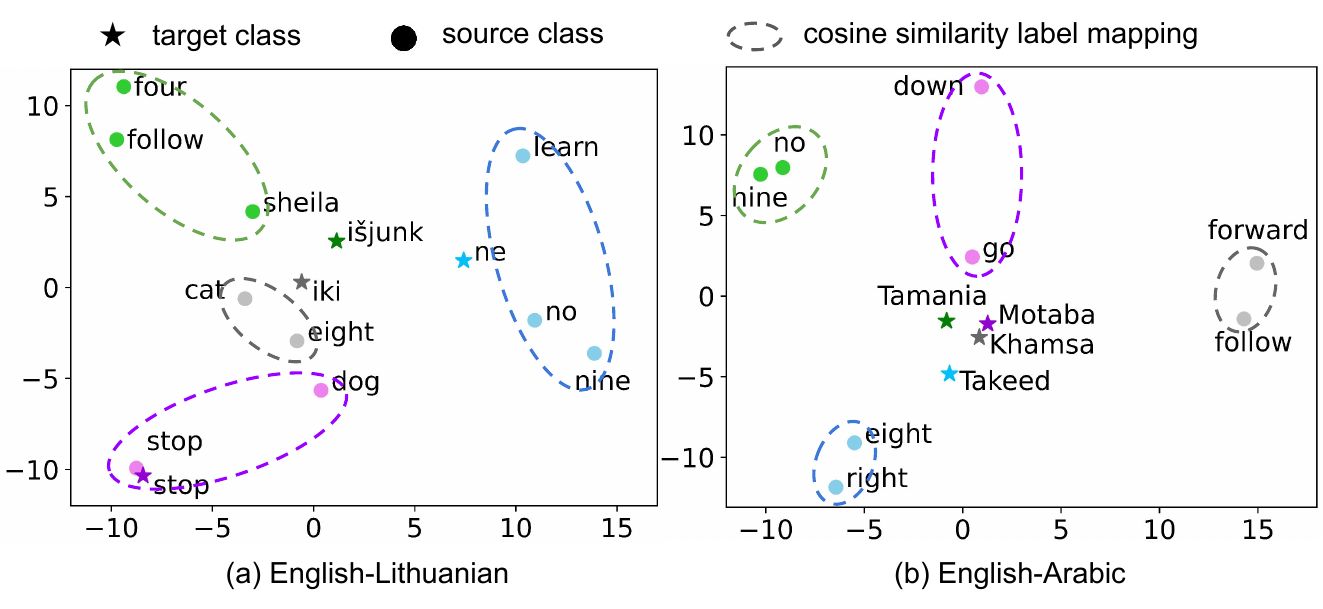}
    \caption{PCA plots of average representations of several source-target pairs for (a) English-Lithuanian and (b) English-Arabic datasets. A target class (star point) is mapped to two or three source classes (circle points) with higher cosine similarity (marked with same colors).}
    \label{fig:mapping}
\end{figure}

As illustrated in Fig.~\ref{fig:illustration}, a label mapping function is adopted to map the probabilities of the source class to that of the target class. The results in~\cite{pmlr-v139-yang21j} show that a many-to-one label mapping strategy (randomly mapping multiple classes from the source task to an arbitrary target class) yields a better performance as compared to the one-to-one mapping strategy. In this study, we attempt to improve the random mapping process with a similarity mapping that considers the relationships of data in the source and target domains. In~\cite{Hu2020}, the structural relationships between acoustic scene classes are explored and utilized to address the domain mismatch issue, while the authors in~\cite{Lample2017} proposed an unsupervised learning method that maps sentences in two different languages into the same latent space for the Machine Translation task. Inspired by the prior art, we propose to investigate the similarity of the labels between the source and target domains and determine the optimal many-to-one label mapping strategy. To compute the similarity, we first feed source and target data to the pretrained AM and calculate the average representations of all classes and compute the cosine similarity between each of them. Based on the similarity of classes, each target class is mapped to two or three source classes in our AR-SCR system.
Fig.~\ref{fig:mapping} (a) and (b), respectively, show the PCA plots of representation vectors for the English-Lithuanian and English-Arabic datasets. Interestingly, we can observe that command words with similar acoustic characteristics are mapped to the same target word. For example, in Fig.~\ref{fig:mapping} (a), the source English source classes "nine, no, learn" are mapped to the target Lithuanian class "ne"; whereas in Fig.~\ref{fig:mapping} (b), the source English classes "right, eight" are mapped to the target Arabic class "Takeed". In our experiment, the AR system with similarity mapping strategy outperforms the one with random mapping~\cite{pmlr-v139-yang21j} by increasing the average testing accuracy by 2.1\%, 13.4\%, 9.2\% on the Arabic, Lithuanian, and Mandarin datasets respectively.

\section{Experiments}
\subsection{Speech Commands Dataset}
We use the Google Speech Commands dataset~\cite{Warden2018} as the source domain data to pretrain our AM. Three low-resource SCR datasets, including the Lithuanian, dysarthric Mandarin, and Arabic speech command datasets, are used as the target domain tasks.
\\
\textbf{Lithuanian Speech Commands:} The content of the Lithuanian speech commands dataset~\cite{Kolesau2020} is created by translating 20 keywords from Google Speech Commands~\cite{Warden2018} into the Lithuanian language. The dataset consists of recordings from 28 speakers, with each speaker pronouncing 20 words on a mobile phone. We follow the setting in~\cite{Kolesau2020} and~\cite{Kolesau2021} and chose 15 target classes: 13 command words, 1 unknown word, and 1 silent class. The resulting dataset consists of 326 recordings for training, 75 for validation, and 88 for testing.
\\
\textbf{Dysarthric Mandarin Speech Commands:} The dataset contains 19 Mandarin commands, each uttered 10 times from 3 dysarthric patients~\cite{Lin2021}, with a 16kHz sampling rate. These 19 commands include 10 action commands and 9 digits, which are designed to allow dysarthric patients to control web browsers via speech. By removing very long recordings, we select 13 short commands in our experiments, and the duration of each command is around one second. We follow the setting in~\cite{Lin2021} to split the whole dataset into 70\% and 30\% to form training and testing tests.
\\ 
\textbf{Arabic Speech Commands:} The Arabic speech commands dataset \cite{Benamer2020} consists of 16 commands, including 6 control words and 10 digits (0 through 9), and each command has 100 samples, amounting to 1600 utterances in total. 40 speakers are involved in preparing the dataset. The speech utterances are recorded at a sampling rate of 48 kHz and then converted into 16 kHz in our experiments. We follow the setting in~\cite{Benamer2020} and split the dataset into 80\% for training and 20\% for testing. In addition, we randomly excerpt 20\% of the training data to form a validation set.

\begin{table}[!t]
\vspace{-0.2cm}
\caption{Testing results (average and std of the accuracy scores) of three conditions (3, 10, and 20 training samples) for the Lithuanian speech commands dataset. The results of the state-of-the-art system ~\cite{Kolesau2020} are also reported for comparison.}
\label{tab:LT}
\begin{adjustbox}{width=0.48\textwidth}
\begin{tabular}{ccccc}
\toprule
\textbf{Limit} & \textbf{System} & \textbf{Avg. Acc. (\%)} & \textbf{Rel. Imp. (\%)} & \textbf{Std.} \\ \midrule \midrule
3     & Baseline      & 33.8             & --                   & $\pm 8.82$ \\
      & AR         & 25.1             & -25.7                &$\mathbf{\pm2.27}$      \\
      & TL         & 52.8             & 56.2                 & $\pm7.69$  \\
      & AR+TL & \textbf{58.9}    & \textbf{72.8}                 &$\pm4.84$  \\ \midrule \midrule
10    & Baseline      & 63.7             & --                   & $\pm5.35$ \\
      & AR         & 34.4             & -46.0                &$\mathbf{\pm4.1}$      \\
      & TL         & 80.7             & 26.7                 & $\pm5.65$ \\
      & AR+TL & \textbf{81.9}      & \textbf{28.6}                 & $\pm4.44$  \\  \midrule \midrule
20    & Baseline      & 70.3             & --                   & $\pm6.8$  \\
      & AR         & 34.6             & -51.4                & $\pm3.0$     \\
      & TL         & 86.8             & 23.5                 & $\pm2.92$ \\
      & AR+TL & \textbf{88.6}    & \textbf{26}             & $\mathbf{\pm1.7}$  \\
\bottomrule
\end{tabular}
\end{adjustbox}
\end{table}

\subsection{Experimental Results}
For each of the three datasets, we compare the SCR results of the four systems, as shown in Fig.~\ref{fig:framework}. The results of the baseline system (denoted as Baseline), stand for an AM trained from scratch on the training set of the three SCR datasets. The results of the systems with AR and TL model adaptations (Fig.~\ref{fig:framework} (b) and (c)) are denoted as AR and TL, respectively. The results of combining the AR and TL methods referred to Fig.~\ref{fig:framework} (d) are denoted as AR+TL. In our preliminary experiments, we tested the performance of several different setups. We have compared the domains for signal reprogramming by directly modifying the input waveforms or modifying the input spectral features. Meanwhile, we evaluated different label mapping techniques, i.e., one-to-one versus many-to-one, as well as random mapping versus similarity mapping. 
In the following discussions, we only report the results of the best setups. For all the experiments in this study, we tested each SCR system 10 times and the average accuracy and standard deviation (std.) values are reported in each table. 

\begin{figure}[!b]
    \centering
    \includegraphics[width=0.8\linewidth]{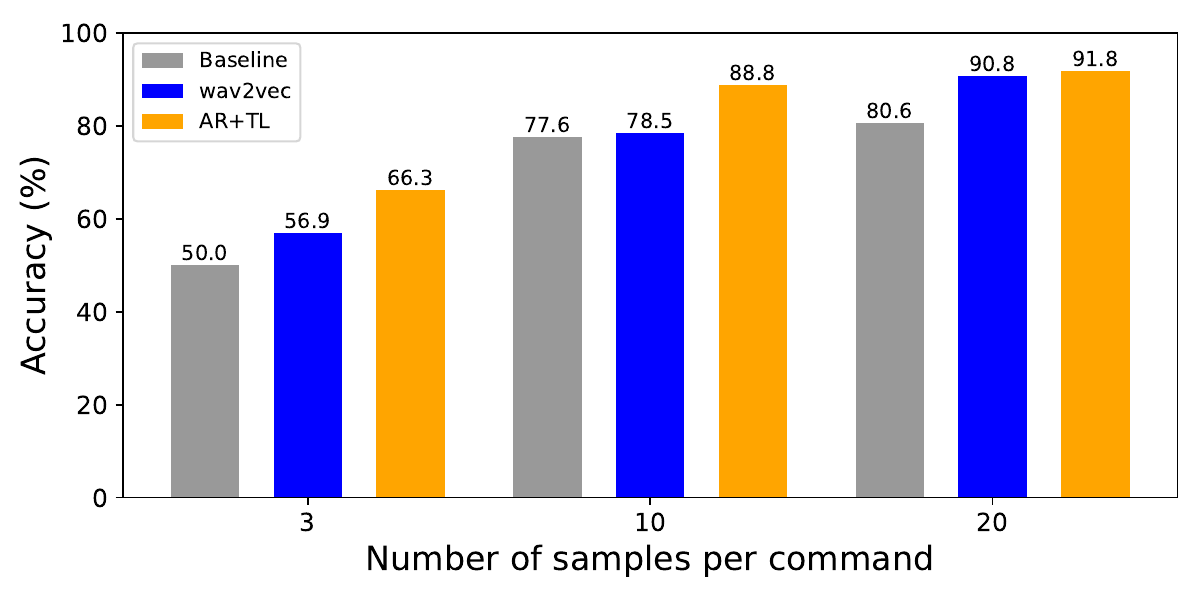}
    \caption{The best-1 accuracy values of Baseline, the state-of-the-art wav2vec model~\cite{Kolesau2020}, and the proposed AR+TL system on the Lithuanian speech command dataset. We follow the same evaluation metrics in~\cite{Kolesau2020} and only reported the best-1 test accuracy values.}
    \label{fig:best-1}
\end{figure}

Table~\ref{tab:LT} lists the testing results of the Lithuanian speech commands dataset. We follow the setting in~\cite{Kolesau2020} to conduct experiments using different amounts of training data and report the results of three conditions: each keyword has 3, 10, and 20 training samples. From Table~\ref{tab:LT}, we observe that TL consistently outperforms Baseline for the three conditions. Notably, although AR alone cannot attain improved performance, AR+TL can yield average accuracy of 58.9\%, 81.9\%, and 88.6\% for 3, 10, and 20 training samples conditions, respectively, which are notably better than TL with the average accuracy of 52.8\%, 80.7\%, and 86.7\%. Moreover, the std values in Table~\ref{tab:LT} show that AR can improve performance stability when combined with the TL technique. Fig.~\ref{fig:best-1} also compares our best AR+TL system with the state-of-the-art SCR system~\cite{Kolesau2020}, where the best-1 accuracy results are reported for the 3, 10, and 20 conditions. From the figure, AR+TL consistently outperforms the state-of-the-art SCR system~\cite{Kolesau2020} on the Lithuanian speech commands dataset. 

\begin{table}[!t]
\caption{Testing results (average and std of the accuracy scores) of the dysarthric Mandarin dataset.}
\label{tab:ZH}
\begin{adjustbox}{width=0.48\textwidth}
\begin{tabular}{ccccc}
\toprule
\textbf{System} & \textbf{Trainable para.} & \textbf{Avg. Acc. (\%)} & \textbf{Rel. Imp. (\%)} & \textbf{Std.} \\ \midrule \midrule
Baseline        & 200.4k                & 64.0            & --                      & $\pm16.43$ \\ 
AR           & 16k                   & 33.2            & -48.12                  & $\pm2.88$ \\ 
TL           & 200.4k                & 78.1            & 22.03                   & $\pm3.18$ \\ 
AR+TL   & 217.2k                & \textbf{82.3}   & \textbf{28.59}                   & $\mathbf{\pm2.56}$ \\ 
\bottomrule
\end{tabular}
\end{adjustbox}
\end{table} 

Table~\ref{tab:ZH} shows the results of the dysarthric Mandarin dataset. A similar observation can be obtained as those from Table~\ref{tab:LT}. First, TL yields an improved performance as compared to Baseline. Next, although AR alone can not provide better results than Baseline, AR+TL achieves the best performance among the four systems. As compared to Baseline, AR+TL yields notable improvements of 28.59\% (from 64.0 to 82.3). Moreover, the low std value of $\pm2.56$ suggests that AR can improve the stability of the SCR systems.

We also conducted experiments on the Arabic speech commands dataset. From the testing results, we first note that the three adaptation systems (AR, TL, and AR+TL) all yield improved accuracy rates with lower std as compared to Baseline. Next, we note that AR+TL can achieve the highest accuracy of 98.9\%, which outperforms individual AR (94.8\%) and TL (98.6\%). Based on our literature survey, AR+TL also outperforms the state-of-the-art SCR system using LSTM~\cite{Benamer2020} (98.13\%) on this Arabic speech commands dataset.

\subsection{Ablation Study: Combining Adversarial Reprogramming with Data Augmentation}

As mentioned in Section~\ref{sec:system}, AR could be easily combined with TL and achieve the highest performance. It is thus very natural to assume that we can combine AR with other low-resource training techniques and further improve the performance of the whole system. To verify whether AR has such flexibility, we combine AR with Data Augmentation (Aug), another standard approach when training DL-based systems with a very limited amount of data. In our experiments, we utilize SpecAugment~\cite{Park2019}, a common data augmentation method in the speech processing field which randomly masks input features by zero along both time and frequency axes. We dynamically perform augmentation to generate additional data during training. We compare the results by combining data augmentation with the Baseline, TL, and AR+TL systems mentioned above. We respectively denoted the three systems as Baseline+Aug, TL+Aug, and AR+TL+Aug.

Table~\ref{tab:DA} shows the testing results of the Lithuanian Speech command dataset when each keyword has 3 or 10 training samples. Compared with the results in Table~\ref{tab:LT}, data augmentation can improve and stabilize the performance of the AR and TL systems by a large margin. Similar results can be observed as those from Table~\ref{tab:LT}. First, TL+Aug improves the performance by 66.8\% and 33.0\% as compared to Baseline+Aug. Then, AR+TL+Aug again achieves the highest average accuracy among the three systems, with 70.2\% and 87.4\% accuracy, which is notably better than TL+Aug (65.9\% and 81.7\%) and Baseline+Aug (39.5\% and 61.2\%), under the conditions of 3 and 10 training samples. Moreover, the lower std values also show that AR combined with data augmentation can improve the stability of the systems. The results show that AR could be successfully combined with data augmentation schemes and further improves the SCR system under low-resource scenarios.  

\begin{table}[!t]
\caption{Testing results (average and std of the accuracy scores) of two conditions (3, and 10 training samples) for the Lithuanian speech commands dataset. }
\label{tab:DA}
\begin{adjustbox}{width=0.48\textwidth}
\begin{tabular}{ccccc}
\toprule
\textbf{Limit} & \textbf{System} & \textbf{Avg. Acc. (\%)} & \textbf{Rel. Imp. (\%)} & \textbf{Std.} \\ \midrule \midrule
3     & Baseline+Aug    & 39.5 & - & $\pm8.65$\\
      & TL+Aug          & 65.9          &   66.8        & $\pm4.52$  \\
      & AR+TL+Aug & \textbf{70.2}       &   \textbf{77.7}        & $\pm\textbf{2.72}$  \\ \midrule \midrule
10    & Baseline+Aug   & 65.3 & - & $\pm8.80$\\
      & TL+Aug         & 81.7           &   25.1        & $\pm2.64$ \\
      & AR+TL+Aug  & \textbf{87.4}      &   \textbf{33.8}        & $\pm\textbf{2.63}$  \\
\bottomrule
\end{tabular}
\end{adjustbox}
\end{table}

\section{Conclusion}
In this paper, we introduce an AR approach to establish an SCR system with a very limited amount of training data. Experimental results show that the proposed AR-SCR system can yield better performance as compared with state-of-the-art SCR methods on the Lithuanian and Arabic speech command datasets~\cite{Benamer2020, Kolesau2020}. The results also demonstrate that AR can effectively improve the accuracy over the baseline system with few trainable parameters on the Arabic dataset. Meanwhile, we also show AR has great flexibility to combine with different low-resource training techniques, such as transfer learning and data augmentation. In our future work, we will explore to further improving the AR-SCR performance by combining it with self-supervised learning methods. Meanwhile, we will investigate the application of the proposed AR approach to improving the model 
safety for both classification and regression tasks. The codes used in this study have been released at~\url{https://github.com/dodohow1011/SpeechAdvReprogram}.

\clearpage

\begin{spacing}{1.0}
\footnotesize
\bibliographystyle{IEEEbib}
\bibliography{refs}

\end{spacing}

\end{document}